# INNOVATION ECOSYSTEMS THEORY REVISITED: THE CASE OF ARTIFICIAL INTELLIGENCE IN CHINA

## Authors

Arenal, Alberto. Universidad Politécnica de Madrid, Spain

Armuña, Cristina. Universidad Politécnica de Madrid, Spain

Feijoo, Claudio. Universidad Politécnica de Madrid, Spain

Ramos, Sergio. Universidad Nacional de Educación a Distancia (UNED), Spain

Xu, Zimu. Cranfield University Cranfield School of Management, UK

Moreno, Ana María. Universidad Politécnica de Madrid, Spain

## Abstract

Beyond the mainstream discussion on the key role of China in the global AI landscape, the knowledge about the real performance and future perspectives of the AI ecosystem in China is still limited. This paper evaluates the status and prospects of China's AI innovation ecosystem by developing a Triple Helix framework particularized for this case. Based on an in-depth qualitative study and on interviews with experts, the analysis section summarizes the way in which the AI innovation ecosystem in China is being built, which are the key features of the three spheres of the Triple Helix -governments, industry and academic/research institutions- as well as the dynamic context of the ecosystem through the identification of main aspects related to the flows of skills, knowledge and funding and the interactions among them. Using this approach, the discussion section illustrates the specificities of the AI innovation ecosystem in China, its strengths and its gaps, and which are its prospects. Overall, this revisited ecosystem approach permits the authors to address the complexity of emerging environments of innovation to draw meaningful conclusions which are not possible with mere observation. The results show how a favourable context, the broad adoption rate and the competition for talent and capital among regional-specialized clusters are boosting the advance of AI in China, mainly in the business to customer arena. Finally, the paper highlights the challenges ahead in the current implementation of the ecosystem that will largely determine the potential global leadership of China in this domain.

## 1. Introduction

'By 2020, they will have caught up. By 2025, they will be better than us. By 2030, they will dominate the industries of AI.' It was November 2017, and Eric Schmidt, CEO of Alphabet, 'set alarm bells ringing in America' about the relevance of China to artificial intelligence (AI) in the medium and longer term. After some decades involving cycles of excitement and hype but most of the time considerable disappointment, AI has definitely evolved from being just a technological fancy trend to become an international field of competition, and the major role of China is beyond any doubt.

This new paradigm has been possible thanks to the consolidation of several key technological enablers.[1] Despite the existence of important limitations to AI today, for example it is very narrow – domain specific – and does not display the general-purpose reasoning capability of humans (Kelly, 2017), experts and analysts refer to it as a new industrial revolution in its capacity to change economy and society. It is poised to change not only the way in which we think about productivity or our relationship with our environment but also elements of national power. Just as past industrial revolutions transferred power to the more industrialized nations, AI can be a similar game changer at the international level (Bughin, Seong, Manyika, & Joshi, 2018).

In this context, several nations have specific AI strategies, and policy makers from around the world (NESTA, 2019) and international organizations such as the OECD[2] have demonstrated great interest in supporting a broad adoption of AI while managing the potential risks and challenges, balancing innovations and social protection. This global trend is largely led by the two largest technology players, the US and China, two superpowers competing for AI world hegemony (Renda, 2019), with the EU, South Korea, and Japan as next, but minor, contenders.

In particular, China presents some particularities that direct even greater attention to AI. At the macroeconomic level, China can no longer depend solely on capital and labour growth to drive the desired levels of sustainable economic growth. In fact, there has been a recent shift towards innovation as a panacea for increasing both product quality and productivity and, on this basis, expanding the internal consumption and thus decreasing the dependence on external markets whilst augmenting the economy's added value.[3] In this context, China is already a leading global force in the digital economy. In fact, digital technologies, particularly AI, are expected to lead to more efficient products and services, achieving large productivity gains and a more dynamic economy, with businesses able to compete globally and even export *Made in China* digital business models and solutions (Woetzel et al., 2017). It has also been argued that AI could contribute to overcoming the physical limitations of capital and labour and open up new sources of value and growth (Purdy, Qiu, & Chen, 2017).

Nevertheless, there are other interests beyond mere economics. China's recent role as a major global power demands reforms to adapt the established international multilateral system to its new needs. Schemes such as the Belt and Road Initiative or institutions such as the Asian Infrastructures Investment Bank can therefore be seen as instruments through which China can shape a geopolitical environment that is more favourable to its interests (Delage, 2017). Digital technologies – and again AI in particular – are essential components of this strategy, showcasing a new, wholly domestic, independent, and cutting-edge industry and a technology-based approach to economics and politics with a different slant from the West. The strict application of national laws to cyberspace or the usage of AI-related technologies to contribute to social harmony are prominent examples of this differentiated approach in China context which in turn demands for an in-depth investigation.

[1] (i) The exponential growth of data – coming mostly from the Internet of Things (IoT) – that can be used to train learning machines, combined with (ii) more powerful computer processing capabilities that can be used for deep neural networks and other learning techniques and with (iii) advances in the algorithms that can make machines very effective at solving a variety of problems across industries.

[2] Information about the OECD initiatives is available at http://www.oecd.org/going-digital/ai/

[3] It is unclear how this productivity increase can be reconciled with high rates of employment while keeping the amount of debt under control and maintaining economic growth. See, for instance, Pettis (2018) for a critique.

Beyond these high-level insights into the Chinese success factors and the general agreement considering China as a paradigm when mapping global approaches to AI, the research works analysing the development of AI in China are limited. In particular, the concept of innovation ecosystem has been popularly used to study the environment that supports innovation activities which AI has become a non-negligible part. However, to the authors' knowledge, no previous studies have been undertaken from an ecosystem perspective. As a result, there is still a gap between the mainstream discussion around the role of China in AI and the knowledge about who the key stakeholders are and which of their relationships configure the AI landscape in China.

Therefore, the main aim of this paper is to provide a holistic overview of the current status, a description of the particularities, and -to the extent it is possible- an anticipatory insight into the AI industry in China by means of a survey of the existing deployments and possible future developments. The study adopts an innovation ecosystem perspective to include and examine all the key players that could have an influence on its evolution. Based on the foundations of the innovation ecosystem theory, the authors propose a revisited asymmetric Triple Helix (ATH) model for the qualitative analysis of the AI innovation ecosystem. The ATH model in this paper is an evolution of that developed by Cai (2013), adding insights from the particular institutional logics in China. In particular, the paper examines the central position of the government (not only at the national but also at the regional and local levels); the role of large companies and start-ups in commercializing the new business models; the way in which public and private venture capital is fuelling growth; and the main initiatives of universities and research centres to produce new knowledge and train people. A secondary objective is to further enhance the understanding of how governmental initiatives work as innovation engines in the industry, especially in planned economies, using the example of AI in China. Thus, the paper aims to both expanding the knowledge about the AI ecosystem in China and contributing to the innovation ecosystems theory. This paper also stresses how a qualitative approach is useful to tackle the complexity in emerging environments of innovation and allow the extraction of insights and gaps that are not directly available from a quantitative analysis of the ecosystem. In addition, the paper shows that qualitative methods are an alternative and/or a complement to quantitative analysis in those ecosystems in which the availability of up-to-date statistics is limited, as in the case of a rapidly changing environment such as AI in China.

This paper has the following structure. After this introduction to the relevance of studying the developments and particularities of AI in China, the next section establishes the research framework of the paper by reviewing the notion, roots, and evolution of the innovation ecosystem approach. Drawing from this theoretical review the paper describes the particularities of the development of AI in China, the Triple Helix as a useful model to study the innovation dynamics in China, and then the specific analytical framework used in the analysis. Then, fourth section details the methods and data used in the analysis and the fifth section presents key insights from the collected evidence, arranged and analysed according to the innovation ecosystem. Finally, the conclusion section closes the paper.

## 2. Literature review: Innovation ecosystems, bringing order to chaos

The concept of an innovation ecosystem has gained increasing relevance since the mid-2000s as a framework that is better suited to emerging industries in which the determinants of the supply and the expectations of the demand are the relevant factors. Though, the origin of the concept is thought to be closely related to two other concepts: business ecosystem[4] and innovation system originated by Moore (1993) and Lundvall (1985) respectively.

[4] The term 'ecosystem' was originally coined by the English botanist Arthur Roy Clapham in 1930, but the authentic pioneer in the use of the concept in the science of ecology was Arthur George Tansley (1935), who introduced it as the basic unit of nature linking living organisms and their habitat (Willis, 1997). Nowadays, one of the most commonly accepted definitions is that adopted by the United Nations (1992) during the convention on Biological Diversity, during which an 'ecosystem' was defined as 'a dynamic complex of

Moore (1993) introduced the ecosystem analogy in business field and outlined that companies cannot be viewed as a member of a single industry but as part of an ecosystem in which customer needs are fulfilled by the cooperative and competitive interaction among companies belonging to different sectors. The success of a company is linked not only to its own strength but also to the relations with the rest of the companies in the network (Hakansson & Snehota, 1995). Therefore, the key feature of any ecosystem is co-evolution. Later, Moore delved deeper into the definition of a business ecosystem, reinforcing the essential role of evolving interaction between individuals and organizations in the performance of the companies and the ecosystem itself (Moore, 1996). After Moore, the influence of business ecosystems on companies' health was broadly developed in the field of strategic management, including the development of conceptual frameworks to assess the situation of a company (Iansiti & Levien, 2004). Broader approaches consider that the ecosystem is formed by interacting 'players', including companies and institutions that provide knowledge and resources and set 'rules of the game' that evolve over time (Feijoo, Gómez-Barroso, Aguado, & Ramos, 2012; Fransman, 2014). The business ecosystem concept has later influenced the emergence of several other concepts (e.g. innovation and entrepreneurial ecosystems) that are beyond firm level and involve multiple entities and stakeholders (Dedehayir et al., 2018; Gomes et al., 2018; Z. Xu & Maas, 2019).

The rapid rising recognition of innovation ecosystem is also thought to be related to the concept of 'systems of innovation' or 'innovation systems' from the late 1980s and the 1990s (B.-Å. Lundvall, 1988). Different institutional models were proposed to characterize innovation processes at national (Freeman, 1995; B.-Å. Lundvall, 1992; Nelson, 1993) or regional (Asheim & Isaksen, 2002; Cooke, Uranga, & Etxebarria, 1997) levels. These models share the focus on structural aspects: they emphasize the interactions and linkages among actors and the importance of agglomeration and geographical placement to facilitate innovation and related commercialisation activities. However, they underestimate the relevance of the dynamic nature of innovation and not explain the relationships between innovators, their innovative activities and the overall environment (Mercan & Götkas, 2011). To address this complexity, the innovation ecosystem approach emerges as a response (Jucevičius & Grumadaitė, 2014). Also drawing from the business ecosystem literature inspired by ecology theories, it highlights that an ecosystem is more than a sum of its parts (Vasconcelos Gomes, Figueiredo Facin, Salerno, & Kazuo Ikenami, 2016).

Innovation 'happens' as a result of the interdependency between players, processes, and their interactions (Fransman, 2018, p. 8). Digitalization leads to a new and more complex logic of innovation involving multiple and heterogeneous stakeholders and their evolving relationships, which could be conceptualized by drawing on the metaphor of a biological ecosystem (Kolloch & Dellermann, 2018). Though there is no universal agreed definition, broadly speaking, an innovation ecosystem parallels the environmental concept, in which interrelated elements strive for equilibrium (Jackson, 2011). In innovation ecosystems, not only is innovation considered as a desirable output, but individuals perform an essential role in the ecosystem dynamics. Since the two initial seminal contributions that disseminated the concept (Adner, 2006; Adner & Kapoor, 2010), the academic discourse has rapidly evolved from 'systems of innovation' towards 'innovation ecosystems' and 'entrepreneurial ecosystems'[5] and other ecosystem-based paradigms.[6] In fact, the academic and industry

---

plant, animal and micro-organism communities and their non-living environment interacting as a functional unit'. This definition highlights the interplays between living organisms themselves and their environment as the key characteristic. One important consequence is that the scale and the situation of the ecosystem are relevant but not decisive.

[5] The first mentions of the term 'entrepreneurial ecosystem' appeared in the work of scholars such as Boyd Cohen (2006) and Paul Bloom and Gregory Dees (2008). However, it only gained popularity after Daniel Isenberg's article 'The big idea: How to start an entrepreneurial revolution' (Isenberg, 2010). In that article, Isenberg defined the entrepreneurship ecosystem as a set of individual elements that, appropriately combined and nurtured, can result in a successful milieu for innovation. Since then, the entrepreneurial ecosystem has become an increasingly widespread concept, and a number of academic papers have defined and referred to it; see, among others, Acs, Estrin, Mickiewicz, and Szerb (2014), Auerswald (2015), Autio and Levie (2017), Malecki (2011, 2018), Mason and Brown (2014), Spigel (2016, 2017), Spigel and Harrison (2018), Stam (2014, 2015), or Vogel (2013). The concept has also

literature is now ripe with studies relating innovative and entrepreneurial activities to the notion of ecosystems (Brown & Mason, 2017; Scaringella & Radziwon, 2018).

It is fair to say that the increasing adoption of the ecosystem approach both by academics and by practitioners has also generated discussion and controversies, mainly around the definition of the concept and the doubts about the value added with respect to the 'system' antecedent (Oh, Phillips, Park, & Lee, 2016; Ritala & Almpanopoulou, 2017).

Often sharing same examples (e.g. Silicon Valley), it is worth noting that the innovation and entrepreneurial ecosystems are two closely linked but different concepts (Xu and Maas, 2019). In fact, despite the nuances between them, it is possible to observe some key shared features regarding the complex interdependency among actors and the need for a holistic approach to analyse the evolution of the system and to evaluate its performance. From the beginning, the innovation ecosystem approach has mainly been based on the premise that the user is beyond doubt, so the focus has mainly been on the supply side (Scaringella and Radziwon, 2018), in contrast to the entrepreneurial ecosystem, which explicitly includes a demand side. In addition, both innovation and entrepreneurial ecosystems have been studied largely at local or regional levels and sometimes national level. Moreover, an essential part of both ecosystems are the dynamic interactions and relationships between different stakeholders in pursing value. Furthermore, innovation is a key element in the entrepreneurial process (Drucker, 1985) and commercialisation, which often involves entrepreneurial activities, is also essential for innovation ecosystems. In innovation ecosystems, large companies (emergent and/or incumbents) have dual roles as both suppliers and demanders within the innovation process, making them a key bridge between entrepreneurial and innovation ecosystems (Autio & Cao, 2019). The concept of an entrepreneurial ecosystem is commonly associated to a certain regional context in which (innovative) entrepreneurial activity flourishes (Mason & Brown, 2014).

The rising interest in innovation and entrepreneurial ecosystems has been accompanied by a number of characterization and evaluative frameworks provided by academics, practitioners, and institutions, which are very different in nature. Among the different studies, scholars such as Carayannis and Campbell (2009) and Xu, Wu, Minshall, and Zhou (2018) have developed frameworks to investigate innovation ecosystems.[7] The notion of an innovation ecosystem has also received remarkable contributions from practitioners (Hwang, 2013; Morrison, 2013) and international institutions such as the World Economic Forum (2017) and the OECD[8] (n.d.).

An ecosystem is typically modelled as a structure resulting from the interaction between various industry and innovation actors or stakeholders. The most relevant of these actors are businesses – big companies but also small companies, start-ups and entrepreneurs, financial markets, universities and research-related

---

received remarkable contributions from practitioners such as Brad Feld (2012) and international institutions such as the World Economic Forum (2011, 2014).

[6] For example, business ecosystems and knowledge ecosystems (Valkokari, 2015) or platform ecosystems (Jacobides, Cennamo, & Gawer, 2018).

[7] Similarly, entrepreneurial ecosystems have received significant contributions. Among different studies, the Aspen Network of Development Entrepreneurs (ANDE) elaborated one of the most comprehensive reviews about these frameworks. ANDE, in 2013, published a methodological guide conceived with the aim of supporting the labour of practitioners in charge of assessing the entrepreneurial ecosystem in developing areas. Before its completion, ANDE conducted an effective comparative study of nine evaluative frameworks with a broad range of scopes, including educational and research institutions (Babson College and George Mason University), international organizations (the World Bank and the Organisation for Economic Co-operation and Development (OECD)) and institutions (the World Economic Forum), private firms (Koltai and Company), associations (the Council on Competitiveness and GSM Association), and even practitioner approaches (Victor Hwan) (ANDE, 2013).

[8] Along these lines and as a main example, the Innovation Policy Platform (IPP) is an ongoing initiative – developed by the World Bank and the OECD – with the aim of facilitating knowledge on how innovation systems operate. Moreover, it is a space where institutional users from different regions can share good practices. Available at https://www.innovationpolicyplatform.org/

organizations, and non-governmental organizations (NGOs) and governmental institutions (Adner, 2006; Frenkel & Maital, 2014). These economic agents display economic relations but also relate through technological, institutional, sociological, and cultural interactions, and, as it has been widely stated, emerging technologies require new ways of governance and participation of stakeholders (Kuhlmann & Ordóñez-Matamoros, 2017; Misuraca, Broster, & Centeno, 2012).

## 3. AI in China and a revisited Triple Helix approach

### 3.1. Innovation ecosystem and Triple Helix approach

While the value of innovation ecosystem concept is widely recognised, there are still several shortcomings when studying innovation ecosystems (Autio & Llewellyn, 2014) and entrepreneurial ecosystems (Audretsch, Mason, Miles, & O'Connor, 2018), especially due to the inability to provide a structured analysis of these interactions among the stakeholders within the ecosystem and their dynamics and to depict the causal path to enhance innovative and/or entrepreneurial activities. In addressing this limitation, Triple Helix (TH) model (Etzkowitz and Leydesdorff, 2000) emerges as one of the most widely adopted and cited frameworks to study the supporting environment and characterize the relationships among the main stakeholders of innovation ecosystems (Chinta & Sussan, 2018; Pique, Berbegal-Mirabent, & Etzkowitz, 2018).

The TH model of innovation was first set with the seminal publication of 'The Triple Helix, University–Industry–Government relations: A laboratory for knowledge-based economic development' (Etzkowitz & Leydesdorff, 1995). The aim of this approach was initially to depict the evolving interactions among the three key institutions in the innovation process within the knowledge-based paradigm: universities, industry, and government. It provides a suitable framework to study the dynamics within the ecosystem in terms of the evolving role and relationships of the main stakeholders, being applied in different studies that examine the conditions of a certain region to promote innovation (Lawton Smith & Leydesdorff, 2014; Natário, Pedro Almeida Couto, & Fernandes Roque de Almeida, 2012). According to Dzisah and Etzkowitz (2009), the dynamic of the TH is based on three basic elements:

> '(1) the prominent role of universities in innovation, on par with companies and the government in a society based on knowledge;
>
> (2) the collaborative relationship between the three core institutional spheres; and
>
> (3) the helices taking the roles of others.'

Therefore, the Triple Helix model can be seen as an explanatory tool to both depict an existing ecosystem and the main players and roles within, and address additional insights about the evolution of the ecosystem through the study of their activity and interactions. Further configurations[9] of the model have been proposed since then and studies were largely based in Western economies' cases (Cai, 2014).

### 3.2. Triple Helix Model in China context

However, based on centralism,[10] China presents key socio-economic particularities that should not be neglected from an ecosystem perspective. For instance, it has commonly been accepted that the 'border' of an ecosystem is not limited to national/regional contexts (Tsujimoto, Kajikawa, Tomita, & Matsumoto, 2018). But this is not the case in China, where it is possible to establish national boundaries. As the main example, the combination of a huge internal market[11] and the barriers to external competition[12] is favouring the rapid development of the domestic Internet companies by providing them access to this internal market almost exclusively. It is also worthy to emphasize on the key role of the governments in China, not only as legislators and regulators, but also providing data, venture capital, structuring human capital, choosing winners and purchasing AI solutions. And this process happens top down from national to regional and local governments, but also leaves room for experimentation at the regional and local levels as it will be later detailed. As stated by President Xi, China '… should unswervingly follow an independent innovation path featuring Chinese characteristics, stick to the guiding principles of independent innovation, leap-frogging development in key sectors and development supported by science and technology and oriented towards the future' (Xi, 2014).

[9] In particular, the Quadruple Helix, including the 'media-based and culture-based public' and the 'civil society'. as the fourth helix (Carayannis & Campbell, 2009); and the Quintuple Helix, including the 'natural environments of society' (Carayannis, Barth, & Campbell, 2012) [Carayannis, Barth, Campbell 2012].

[10] China's constitution defines its political system as a 'people's democratic dictatorship' aiming to develop a 'socialist market economy'. Available at http://www.npc.gov.cn/englishnpc/Constitution/2007-11/15/content_1372962.htm

While a number of studies have applied the TH model in China context (e.g. Lu and Etzkowitz, 2008; Zhang, Zhou, Porter, Gomila, and Yan, 2014; Zhou, 2008; Jongwanich, Kohpaiboon, and Yang, 2014), the particularities subsequently demanded for a revised Triple Helix model to reflect China's context (Cai, 2014). As Yuzhuo Cai summarized, 'the Triple Helix is indeed the solution for China, but to achieve the end the road is hard due mainly to some unique characteristics of the Chinese context' (Cai, 2014, p. 3). These differences refer not only to the economic structure and organization issues but also broad aspects such as the political system, culture, and social norms. Derived from those reflections and drawing from Etzkowitz's approach (2008),[13] Cai suggested a specific TH model by adding insights from the particular institutional logics in China. The result is a 'delayed government-led model' that represents an evolution of the previous characterization of the TH relationship between university, industry, and government in China as a statist model (Etzkowitz & Zhou, 2006).

Cai and Liu's (2014) TH model is primarily based on the case study of Tongji University Creative Cluster in Shanghai, China. In their work, the initial interactions took place between the university and the surrounding industry (mainly university spin-offs) and were spontaneous. While such university-industry collaborations were encouraged by national policies, governments at different levels were not originally involved. As the cluster grew and conditions changed, the district government got involved as a partner through reflective control. In this second stage, the relationships between university, industry and district government were much like the imbricating Triple Helix model where the three parties 'overlap and collaborate with each other' (Cai and Liu, 2015. P.18). As the cluster's economic and social impact continue to grow, central and municipal government started to get involved. During this stage, with various sources of support provided, central government also incorporated and linked the cluster into the national innovation system. The TH has then evolved to become government-led where different levels of government at central, municipal and district levels took charge of the overall development of the university and industry cluster at particular stages. In a final stage, though the district and municipal governments are still working together with the university, they have also started to take the more traditional role of regulators.

In summary, the government-led model takes into account the dynamic relations among university, industry and government at different levels and provides a suitable explanation on how bottom-up initiatives could be integrated with top-down plans. Thus, the government-led conceptual model proposed by Cai (2013) and Cai

[11] China is the world's most populous country, representing almost 20% of the total world population.

[12] The main example is the Great Firewall, a subproject within the Golden Shield Project. It consists of a range of legislative and technical initiatives aiming to regulate the Internet at the domestic level in China and keep control of China cyberspace. It implies blocking the access to a number of foreign websites and as a consequence a slowdown of cross-border Internet traffic.

[13] According to Etzkowitz (2008), the triple helix is configured on three levels, the two first approaches being preliminary stages for the real innovation helix:

1. In the first stage or *statist model*, the government rules the relationships between academia and industry. Common examples are the traditional approach of countries in LATAM, the URSS, or even France.
2. In the second stage or *laissez-faire* model, the three actors are relatively independent and industry is the driving force. Sweden and the US are common examples of this approach.
3. In the third stage or *balanced* model, which is an evolution from the two previous approaches with no specific driving force among industry, the government, and academia, all three contribute to the knowledge society.

and Liu (2014) is used as the base reference to propose the analytical model for this paper. This model is further particularized and discussed in the next section.

### 3.3. A revisited Triple Helix model for China's AI ecosystem

As a summary and for the scope of this paper, the innovation ecosystem approach permits to address the complexity of emerging environments of innovation to draw meaningful conclusions that are not directly available from other means. Those are the type of challenges involved in analysing the AI industrial ecosystem in China, an ever-changing and many-sided phenomenon located in a complex but particular geography aiming to dominate some key technological industries. Despite presenting some limitations in providing a structured analysis of the interactions among the main actors of the ecosystem, the TH represents an acknowledged framework to describe the relationships between the government, universities, and industry as the main stakeholders. The authors consider Cai and Liu's (2014) TH model is particularly appropriated to describe the innovation context in China, including its particularities, with some adjustments as described in the next paragraphs.

First, Cai and Liu's (2014) TH model is developed primarily by studying the development of a particular local cluster. However, the development of AI in China carries many singularities which pose questions on whether this model is capable of capturing the complete picture. Firstly, AI is being applied to a whole range of industries (Baccala et al., 2018), and it is better to think about AI not as a discrete type of technology but as a basic enabling technology, like electronics or general-purpose software. Kevin Kelly suggested that, just as electricity empowered and enlivened all sorts of objects, AI will similarly 'cognitize' objects, making them more intelligent and useful (Kelly, 2014). In fact, the impact of AI has been noted to be just as transformative as electricity (Lynch, 2017), but, of course, AI will not only benefit the relationship with objects but will also automatize all types of processes – too complex to tackle previously – either in industry (Cotteleer & Sniderman, 2018) or in service-related economic activities. The ecosystem paradigm has been used to analyse not only firm based-ecosystems but also 'broad socio-technical regimes' and 'industry-crossing economic developments' (Almpanopoulou, Ritala, & Blomqvist, 2019), as is the case of AI.

In addition, while Cai and Liu's (2014) TH model focuses on capturing a regional innovation system, this paper is studying the AI ecosystem in China from a country and industry level. Two particular aspects should be considered to this respect. Firstly, from authors' perspective, the digital success of China is driven by three main factors: (i) a large and young Chinese market, enabling rapid commercialization of digital business models fuelled by strong public and private venture capital funds; (ii) a rich digital ecosystem, expanding quickly beyond a few large companies to emerging start-ups; and (iii) strong government support for companies, universities, and research centres, providing favourable economic and regulatory conditions on three axes: acting as an investor, as a consumer of digital technologies, and as a provider of key data for companies experiencing advantageous conditions. Secondly, although China continues to operate on a centrally planned system,[14] with innovation initiatives following a top-down approach, in many instances – AI in particular -, regional and local governments have the mandate to develop their own innovative activities independently from the rest of the country. This situation is an evolution of the traditional cross-regional competition in economic terms (Li, Li, & Zhang, 2000).

Altogether, the authors argue that both the particularities of AI and of China require a revisited approach to Cai´s government-led TH model able to encompass the novelties of this case. As a result, the framework proposed for China´s AI ecosystem is an asymmetric Triple Helix model (ATH) (Figure 1) in which (i) the central government controls the global context of innovation but provides a certain degree of autonomy to regional, local, and even district governments to conduct their own policy experiments, which could be discarded and/or being extended depending on the results achieved; (ii) the role of the industry is re-examined to highlight larger digital companies, but also the start-ups that commercialize these new business models, and the finance domain (from both public and private points of view) with a particular mention of venture capital; and (iii) that includes the role of universities and other R&D centres, producing new knowledge and training people.

[14] Since 2014, the Chinese national government has launched a series of key national economic, regulatory, and policy initiatives that touch upon AI with the aim of creating a €13 billion AI market in China by 2018 and making China the world's leading AI power by 2030.

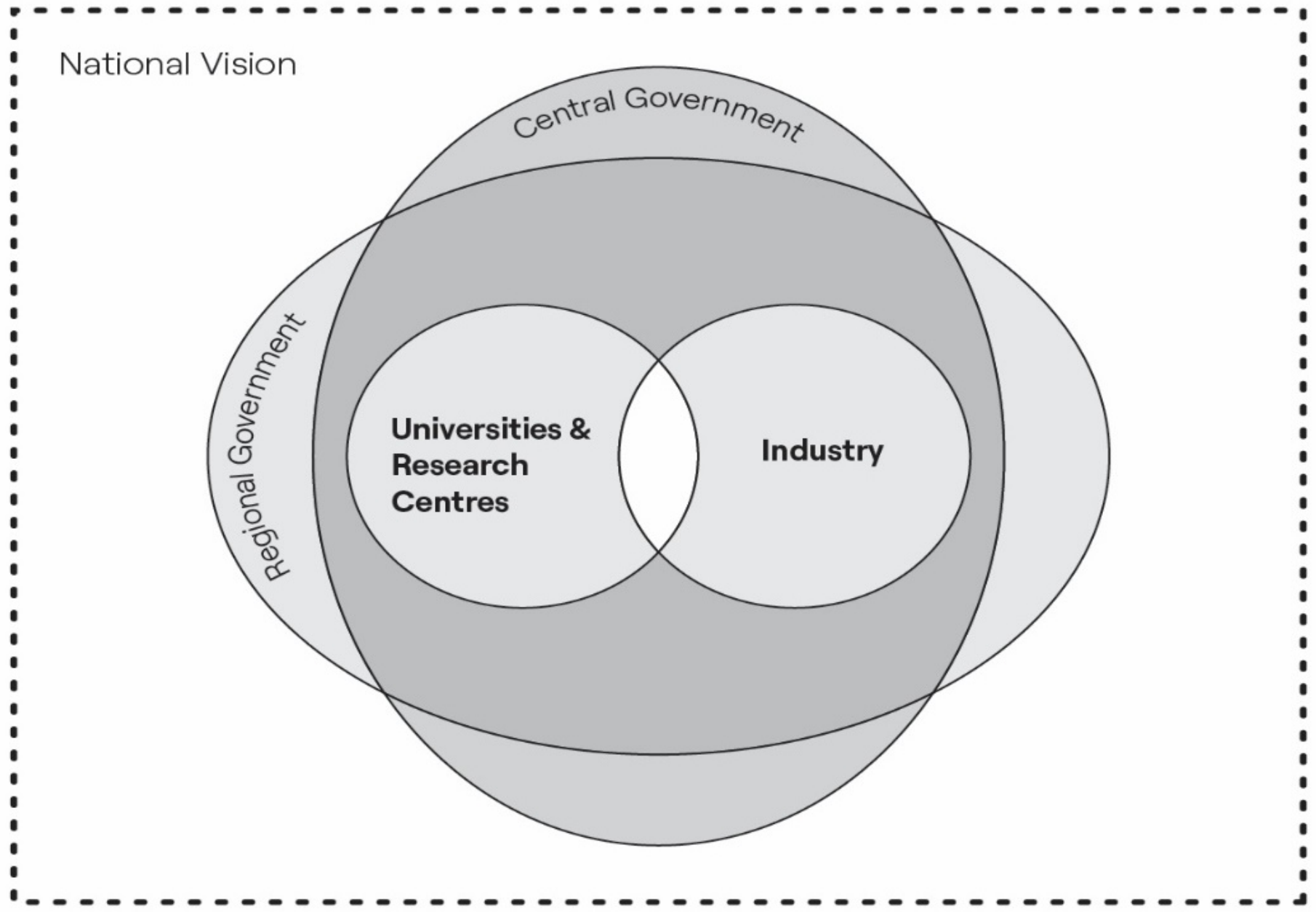

Figure 1 – Asymmetric Triple Helix (ATH) model for the AI ecosystem in China. Source: Own elaboration adapted from Cai (2013) and Cai and Liu (2014).

Therefore, the proposed ATH model differs from Cai and Liu's (2014) model in three main perspectives. The first lies on early government interventions. In the TH model, government intervention comes after the initial university-industry collaborations become successful. However, central government sets up strategic plan and pulls together a range of resources to support the development of AI across the country before any AI clusters started to self-emerge. Secondly, regional and local level governments have certain degree of autonomy in designing their own AI related policies rather than just following largely overpowered by central government. Thirdly, the proposed ATH model has an in-built emphasis on large enterprises, start-ups as well as venture capitals whereas Cai and Liu's (2014) stressed that the one of the characteristics of TH in China lies on the university-owned enterprises which 'integrate several stages of the research, development and commercialisation process into one organisational entity, thereby avoiding a long and complicated negotiating process with other enterprises' (p.20), that it is not necessarily the dominant situation in the AI arena in China.

## 4. Methodology

Descriptive research is conducted to validate the ATH model proposal and the domain analysis developed. The methodology followed combines desk research, fieldwork, and document analysis undertaken during the period 2017– 2018.

After the literature review about the innovation ecosystem approach and the suitability of TH for the particular case of China's AI ecosystem, the next step was building a semi-structured questionnaire to gather expert's vision formed by a total of 32 questions (Annex I), allowing both open and Likert-scale answers. Experts for the survey were selected from the available contacts in the China AI ecosystem, in particular through China

Association of Artificial Intelligence, Chinaccelerator and the main universities devoted to AI in China. The semi-structured questionnaire was distributed through online invitation by email and Wechat using the surveys' provider Typeform between March and June 2018 to 120 experts approximately equally distributed among the three main areas identified in the framework: government, industry, and universities. As some of the experts participated also in validation events during the elaboration of the paper, where reminders were provided on the opportunity of participating in the survey, the response rate of 22% is higher than usual average for online surveys. The respondents (N=26) comprised significant representatives of key AI stakeholders in China, in particular working on AI in governmental bodies or institutions (N=2), industry (N=18: 8 at large companies, 2 at small or medium companies, 6 at start-ups, and 2 venture capital), and universities (N=5). Anonymity was offered, and one participant preferred not to specify his or her working domain. The average age of participants was 38 years, 15% were women and 85% men, and the nationality split indicated that 41% were born in China and 59% were foreign professionals with an average of 9 years of working experience in China.

The questionnaire was structured on four blocks: the AI innovation ecosystem's status, challenges and prospects; the AI ecosystem's leading actors, advantages, and disadvantages; AI's future opportunities and threats in China; and personal background, including expertise in 'AI in China'. Despite the limited sample size, the results contributed to identifying the main particularities of the Chinese AI ecosystem to validate the theoretical model that supported the analysis and the stakeholders' relationships, and primarily serves an explorative and illustrative purpose. The survey was dynamic and used an adaptation of the experiment board methodology (Loon, 2014).

Secondly, taking the proposed model as a reference, the examination of key stakeholders and their relationships was carried out by means of an exhaustive secondary source documental revision, representing the AI initiatives or projects of each of the stakeholders. Textual analysis was performed following a structured revision identifying the name of initiative/project, date, the supporting entity (government institution, organization, company), the AI area target and the description. This desk research gathered a data set as exhaustive as possible from multiple sources as detailed in the following: the national government analysis involved the revision of all the AI policy-related initiatives publicly available during the period 2015–2018 (Table 1); the regional governmental domain was addressed by selecting all the policy initiatives related to AI promoted in TIER1[15] cities (Beijing, Shanghai, Shenzhen, and Guangzhou) and those included in the TIER2 classification highlighted by the survey experts as AI future opportunity areas (Nanjing, Wuhan, Hangzhou, and Chengdu) (Table 2); the industry analysis considered the AI company projects announced by (i) large national companies (Table 3), selecting the four companies designated by the Government as members of the artificial intelligence national team; (ii) large international companies (Table 3) , selecting the four largest US companies with relevant presence in China (Apple, Amazon, Google, and Microsoft); (iii) Chinese unicorns (Table 4), according to industrial rankings (CB Insights, Credit Suisse, 2019); and (iv) start-ups and small companies selected from diverse industry sources. For the university domain analysis, the revision included the top five universities according to the ranking 'Top 60 AI universities of Chinese mainland' (Table 5).

Finally, we presented and discussed the results in two close-door seminars to 27 AI specialists in a roundtable on the AI innovation ecosystem in China within the Global Artificial Intelligence Conference[16] in Beijing, 19–20 May 2018, and a specific event on AI and start-ups organized in Shanghai on 15 May 2018 with the support of Chinaccelerator and SOSV Venture Capital.

[15] China's tiered city system is a hierarchical classification that segments China into four categories considering development factors (gross domestic product (GDP), political administration, population size, development of services, infrastructure, cosmopolitan nature, retail sales, etc.). Although not officially proposed by the Chinese Government, it is commonly used by economists, consultants, and businesses. All TIER1 cities have a GDP over US$300 billion and are directly controlled by the central government; the TIER2 GDP is US$68–299 billion and it includes provincial capital cities and sub-provincial capital cities. https://knowledge.globalwebindex.net/hc/en-us/articles/212633345-Chinese-Tier-Classification.

[16] See https://2018gaitc.caai.cn/en

The result is exploratory in nature and supports the theoretical research on the current AI ecosystem in China that updates previous attempts at characterization to provide a conceptual scheme of analysis in a more systematic way. Following sections include the analysis performed and conclusions.

## 5. Ecosystem analysis and discussion of the AI development in China

The analysis follows the revisited ATH model described above and starts with the review of both the general and specifically-related-to-AI government and policy initiatives since 2014. The analysis aims at the validation of the base hypothesis of this paper on the central role that the institutional bodies -including regions and cities- are playing in the AI ecosystem in China. From there, it continues an investigation of the most relevant AI projects under development in China led by the main –most valued- large companies and start-ups and, in a third step, the review of the status from universities, research centres and government funded research-oriented initiatives. Next, the analysis examines the dynamics of the AI ecosystem, this is, the relationships between the three main helices / stakeholders. For this the paper has selected venture capital (government-industry relationship), human capital (industry-universities relationship), knowledge production (universities-industry relationship), and data availability framework (government-industry relationship). Finally, the results of these dynamics are captured in a specific section on selected case studies.

### 5.1 Government

#### *5.2.1. National initiatives*

Table 1 summarizes all the national AI policy initiatives in China in the period 2015-2018. They are classified by the supporting entity, the position within the AI value chain, and the aim of the initiative.

Table 1 – AI policy-related initiatives at the national level in China (2015-18). Source: Authors' own compilation from desk research.

| Initiative | Launch date | Supporting entity | Position within the AI value chain | Vision/aim |
|---|---|---|---|---|
| Made in China | 20 May 2015 | State Council | Industry | China will become an advanced and prestigious manufacturing power by 2025 |
| Internet + | 4 July 2015 | State Council | Services, infrastructure | Creating an industrial ecology of active innovation, open cooperation, and coordinated development |
| Robot Industry Development Plan | 27 April 2016 | Ministry of Industry and Information Technology, National Development and Reform Commission, Ministry of Finance | Industry | Development of intelligent industrial and service robots |
| 13th Five-Year Plan for National Science and Technology Innovation | 28 July 2016 | State Council | Science and Technology, Industry | AI as a key element of the research and development and innovation strategy |
| Consumer Goods Standards and Quality Improvement Plan | 6 September 2016 | State Council | Apps/content | Commercialization, patentability, and standardization of several technologies (including AI) |
| Cybersecurity Law | 1 June 2017 | Standing Committee of the National People's Congress - State Council | Apps | Establish the legislative framework on cybersecurity |
| Notice on 13th Five-Year Plan for National Strategic Emerging Industries | 19 December 2016 | State Council | Industry | Promote AI application in key economic and social fields |
| Opinions on Promoting Healthy and Orderly | 16 January 2017 | CPC Central Committee | Services | Step up the layout of key AI technologies across |
| Development of Mobile Internet | | General Office – State Council | | mobile industries |
| New Generation AI Development Plan | 8 July 2017 | State Council | Apps/content, Services, Infrastructure, Industry, R&D | Use the first-mover advantage and build a world science and technology powerhouse in the field of AI |
| Three-Year Action Plan to Upgrade Manufacturing Sectors | 29 November 2017 | National Development and Reform Commission | Industry | Stimulate growth in key industries and develop robots with autonomous programming, human–machine collaboration, and dual-arm collaborative robots |
| National-Level Innovation Platform for Intelligent Vehicles | 29 March 2018 | National Development and Reform Commission | Infrastructure, Industry, R&D | Solve the problems and obstacles in the development of intelligent vehicles, attracting key enterprises, encouraging overseas mergers and acquisitions, and enhancing the research and development capabilities of intelligent vehicles |

From the analysis, it is apparent that, in the last five years, a number of strategic initiatives addressing the whole value chain have proliferated (Figure 2), starting with a specific focus on industry development and slowly taking up all the other elements in the value chain.

A summary of the contents of these initiatives includes (i) the support for the involvement of public and publicly aligned companies, (ii) standardization as a strategic thrust, (iii) innovative purchases from governmental bodies, (iv) issuing and enforcement of all type of regulations and regimes – such as intellectual property, (v) addressing the interactions between the industrial–military complex and the civil sectors, (vi) the promotion of human capital, (vii) the availability of financing, and (viii) the facilitation of access to data.

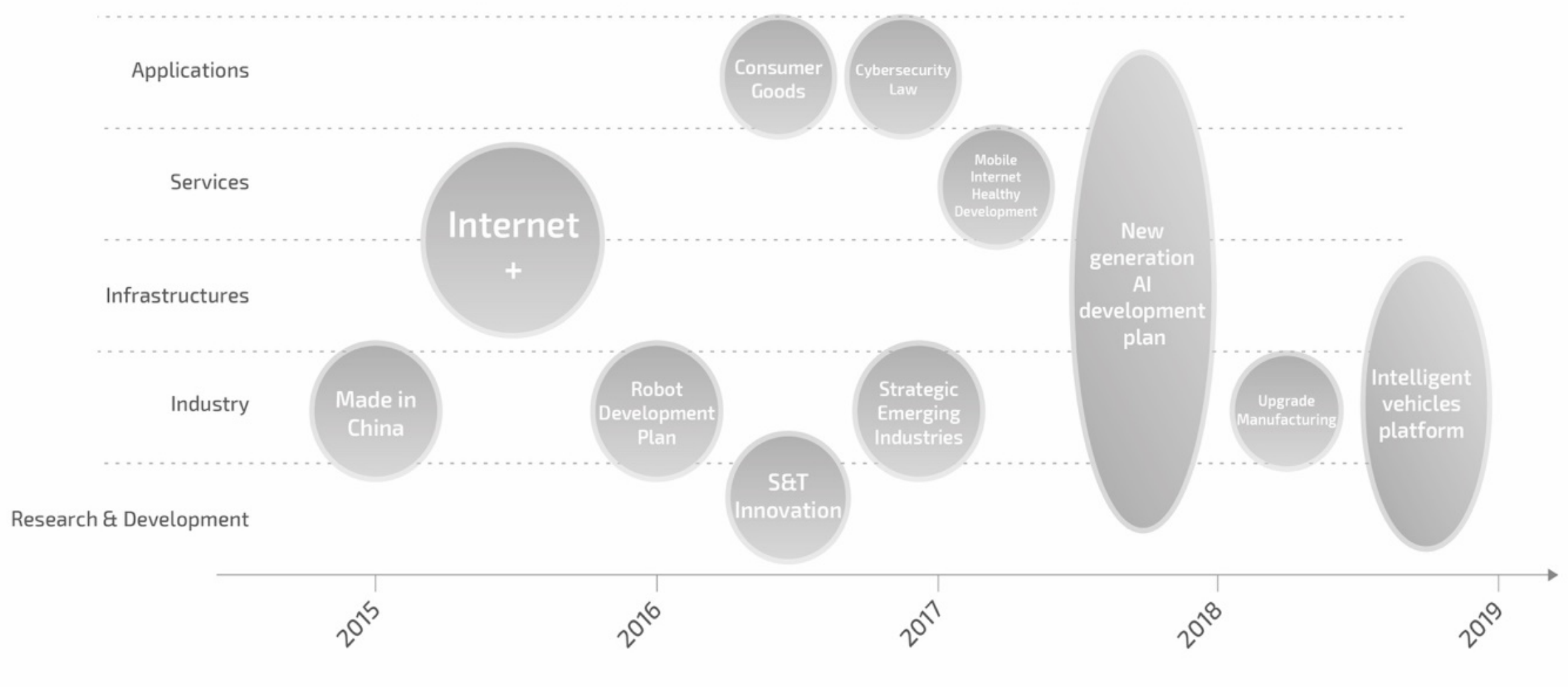

Figure 2 – Key national initiatives that touch upon AI in China (2015–2019). Source: own elaboration.

Missing elements until the end of 2018 were the research in fundamental challenges in AI and the ethics debate on AI. About the former, it is interesting to note that all the main companies in AI in China keep research centres in the US as a means to gain access to potential upcoming disruptive developments. About the latter, the regulatory focus in China is mostly on how to place AI within the scope of the rule of law and not referring to issues related to the transparency and accountability of the algorithms. In this respect, until very recently, an opinion generally held by experts in China was that, with the existing level of technical development, the existing laws and regulations were enough to solve the problems (People's Daily, 2018; Tan, 2018). However, lately some voices have called for new legislation (Cao, 2018) and new regulation is being proposed as of 2019 to curb the excessive gathering of personal data by private companies, and also proposals for self-regulatory AI code of ethics although without much clues on its enforcement.

Arguably, the best examples of the influence of these framework conditions are the multiple and successful solutions based on voice, image and facial recognition systems that are already being deployed, and the different solutions aimed at the so-called social credit -rather trust or reputation- system for citizens. Pending issues in these deployments, however, are lack of standards, issues about the compatibility of solutions, bureaucratic inefficiencies (Godement et al., 2018), and technological limitations.

#### *5.1.2. Regional /local initiatives*

Similarly, there are also many initiatives taking place in China at the regional and local levels, supported by the respective governments, see Table 2 for a compilation of initiatives of TIER1 and most relevant TIER2 cities in the period 2017-2018. In general, China has 17 national-level innovation demonstration zones, which were selected by the State Council and enjoy favourable policies to encourage innovation and regional economic growth. Most of the opportunities lie in the East (Beijing, Shanghai, Hangzhou, Guangzhou and Shenzhen), with the major exception of Chengdu in Sichuan province, although all the main cities have shown commitment to the development of AI. Precisely, one of the main features of the AI innovation ecosystem in China from the policy point of view is the particular type of competition for talent between different locations – or clusters of locations – and particularly the new policies launched by second-tier cities.

From the analysis, it is possible to conclude that policy in China is developed on a top-down basis, but relevant initiatives are also put into practice as experiments at the regional and local levels with the aim of extending those successful projects to other areas or even the whole country. The approach to policy making is radically practical, based on several different – sometimes hardly compatible – strategies, all of them based on a trial-and-error approach by testing new policies in pilot projects. Depending on the outcomes of these tests, policies are abandoned, refined, or rolled out across the country (Stepan & Duckett, 2018).

Table 2 – Main regional/local initiatives related to AI in China (2017-18). Source: authors' compilation.

| | **Beijing** | **Shanghai** | **Shenzhen** | **Nanjing** | **Wuhan** | **Guangzhou** | **Hangzhou** | **Chengdu** |
|---|---|---|---|---|---|---|---|---|
| **Area(s)** | Venture capital, talent, support facilities | Talent, venture capital, support facilities | Talent, venture capital | Talent, critical mass | Talent, critical mass | Support facilities | Support facilities | Talent, critical mass |
| **Vision/ Aim** | AI-oriented funds; enhancing AI talent and support facilities Establishing a demonstrative hub for innovative development in Xionggan (located 100 km southwest) | Promoting talent in emerging technologies; large AI-oriented fund; innovative parks for start-ups | Investment angel funds; large grants for scientific teams (Peacok Initiative), facilitating access to Hong Kong | Special residence visa (hukou) for skilled workers, subsidies to attend job interviews, housing advantages | Residential community to attract graduates through discounts on rentals | Industrial parks related to the manufacturing industry | Incubators partly subsidized by the Government for entrepreneurs (former workers of Alibaba | One week of free accommodation for fresh graduates, promoting quality of life, competing with Hangzhou |

### 5.2 Industry

Following the ATH model, the second step of the analysis addresses the role of the different types of businesses within the ecosystem.

According to the analysis, AI advances in the business-to-business domain seem less developed, mainly due to the difficulties of benefitting from scale economies. Similarly, traditional industries need to work with technology companies to develop specialized AI in a lengthy process with lower returns on investment (Lucas, 2018).

#### *5.2.1 Large companies*

From the analysis, it can be noticed that three different groups of large established companies play a leading role: (i) the big three internet giants in China– Alibaba, Baidu, and Tencent- each with a particular focus on ecommerce, location/navigation and healthcare respectively, and overlaps in payments and social networks; (ii) a second group of large companies partially involved in AI, including the ride-hailing company Didi Chuxing, the on-demand service provider Meituan-Dianping, the mobile handset and network equipment manufacturer Huawei (and also ZTE and Xiaomi, respectively in networks and handsets), and the speech and language recognition firm iFlytek – formerly a unicorn[17]; and (iii) a third and last group of large foreign tech companies – mostly from the US – that have set up research centres and joint ventures in China to position themselves in what they thought could be a promising innovation landscape, such as Apple, Amazon Web Services, Google and Microsoft. No European large company holds any relevant position in the AI innovation ecosystem, to the authors' knowledge. Figure 3 displays the areas of these three groups of large companies in China's AI innovation ecosystem.

As a summary, China's large companies are both particularly supportive of the government initiatives and use the positive perception of consumers to experiment with applications that are unheard of outside China, mainly in areas related to online consumption, security, public safety, public administration and social networks.

In addition, the limited number of top companies in AI in China, as in technological sectors elsewhere, causes investors to tend to bet on two to three dominant players, and these dominant players usually capture all the emergent start-ups to ensure the continuation of their dominant positions, in particular in general consumer-related markets. Related to this growth strategy, another common approach across large companies in AI in China is funding research centres abroad (mainly in the US) in addition to the centres on the Chinese mainland. As mentioned, their aim is to tap into new knowledge and potential disruptive advancements in this domain. At the same time, US companies -Google, Microsoft, Apple- keep research centres in China to tap on knowledgeable human capital and attract developers (and market) to their platforms. Also, it is worth to mention that the hardware and software driving most of China's AI comes from US companies – chips from Qualcomm and Nvidia and software from Google's TensorFlow. Since the US ban – now rescinded – on ZTE and the complicate situation with the Huawei case, all the main Chinese tech companies have accelerated their plans – within the 2025 Made in China policy – to build their own AI chip sets, aiming to achieve a higher level of independence in chip manufacturing or even leap-frogging it towards specialized AI chips, but the current situation is that China's semiconductor industry is still lagging behind.

[17] Lately, Bytedance could well belong to this section (it is still included in the next section on start-ups), not only for its valuation, but also because of its decision on finding its own space disregarding acquisition offers from Tencent. Bytedance uses AI to engage users, among others, with images and short videos (Douyin, Tiktok, Helo, …) and news (Toutiao).

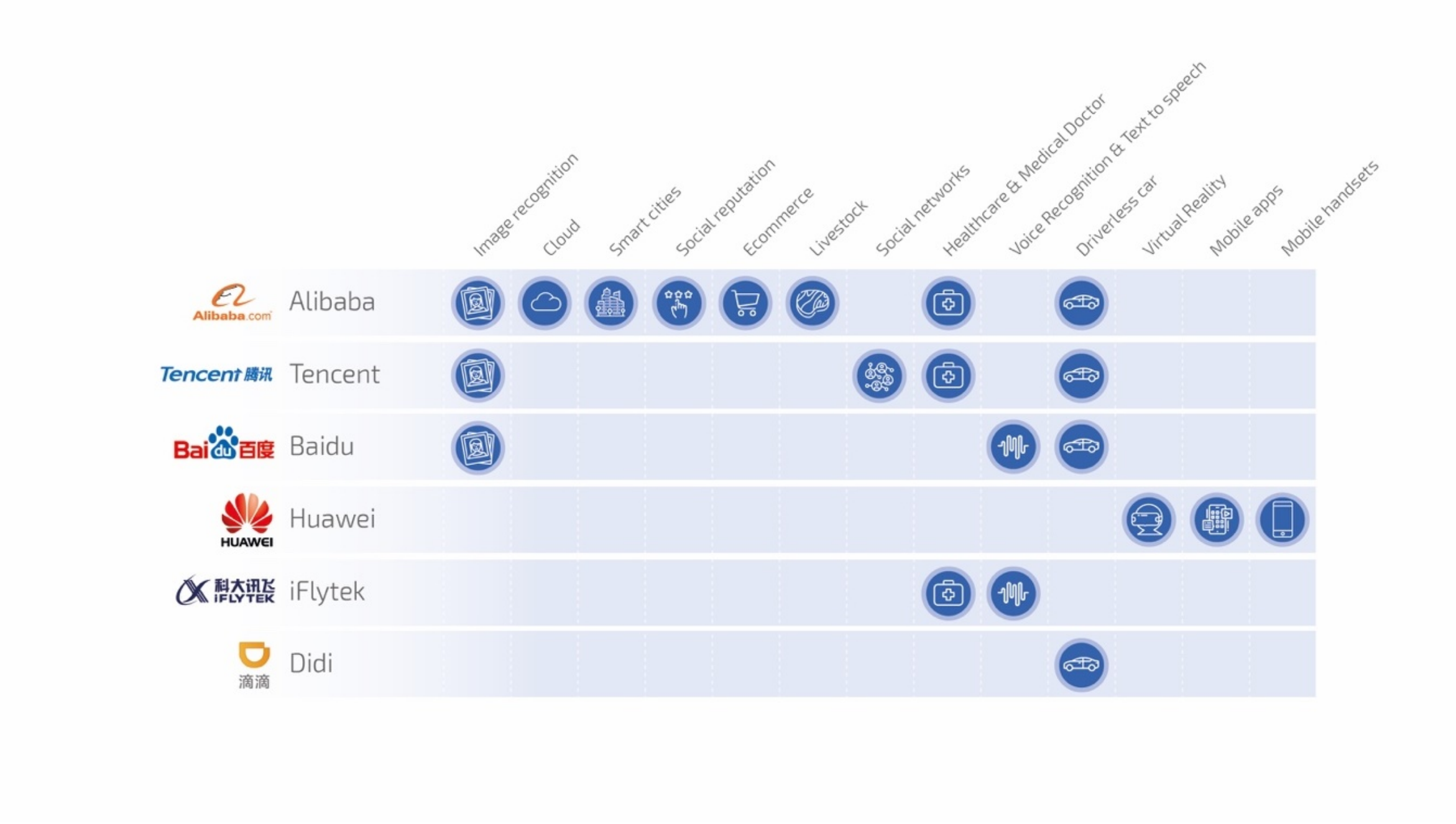

Figure 3 – Areas of specialization of the main large companies in China's AI innovation ecosystem (2017-18). Source: own elaboration.

### *5.2.2 Start-ups*

Table 3 summarizes all the reported AI related start-ups in China that reached unicorn status in 2017-18, including a brief description of the main aspects of the projects that they lead, and their market value.

| Company Name | Headquarters | Aim / Description | Market Value ($bn) |
|---|---|---|---|
| Pony.ai | Freemont (CA) | Autonomous driving | 1.0 |
| Mobvoi | Beijing | Voice recognition, natural language processing, and vertical search technology in-house | 1.0 |
| Tongdun Technology | Hangzhou | Intelligent risk management and decision making | 1.0 |
| iCarbonX | Shenzhen | Personalized health care | 1.0 |
| Momenta | Beijing | AV perception software | 1.0 |
| Unisound | Beijing | Voice technology and AI chip for NLP applications | 1.0 |
| Face++ | Beijing | Image / facial recognition technology | 1.0 |
| 17zuoye | Shanghai | Personalized education platform | 1.0 |
| 4Paradigm | Beijing | Anti-fraud for insurance and banking | 1.2 |
| Horizon Robotics | Beijing | Robotics | 1.5 |
| Yitu Technology | Shanghai | Facial recognition technology and medical imaging and diagnostics | 2.4 |
| Cambricon | Beijing | Neural AI processors | 2.5 |
| Cloudwalk | Guangzhou | Facial recognition technology | 3.3 |
| SenseTime | Hong Kong | Image / facial recognition technology | 4.5 |
| Ubtech Robotics | Shenzhen | Robotics | 5 |
| DJI | Shenzhen | Drones | 15 |
| Bytedance | Beijing | Personalized videos and news curation | 75.0 |

Table 3 – Valuation and headquarters of AI-related unicorns in China as of January 2019. Source: own elaboration based on market analysts' data.

Most of the unicorns are based in Zhongguancun in Beijing (the so-called China's Silicon Valley), though there are others located in Shenzhen (more hardware-related), Shanghai (service-related), Hangzhou (around the Alibaba's ecosystem) or Guangzhou. This is another indication of opportunities being concentrated in some regional/local specialized clusters competing for attractiveness. In the case of AI, although there is considerable competition for talent and the ubication of AI-related start-ups between regions and cities in China, most of the relevant universities, venture capital and national government connections are located in Beijing.

Regarding the specialization domain, the ratio of unicorns based on advanced scientific research capability (health care, biotech, software, robotics, AI, etc.) is still higher in the US than in China as of 2019 in general. Chinese unicorns are more focused on platforms and business serving directly consumers' needs (B2C) and not those of companies (B2B) and use the size of the market as an advantage. Some analysists have remarked that these features reflect, on the one hand, that China's consumer market is large and emerging and, on the other, that China is still in the catch-up phase in terms of science and technology development with respect to the US (Credit Suisse, 2019). From here, according to experts, any consumer-related business will mean in the mid-term that already existing large companies will be interested in this same market, and that the most probable scenario in China is different large companies dominating in each service / consumer-based sector with a clear segmentation between them.

It should also be noted that, although there are no foreign-owned AI unicorns in China, there is considerable activity of non-fully Chinese AI-related start-ups in China. Sometimes they use the traction of the Chinese market – a very tough market but with easier access to medium-sized companies – to gain additional momentum and return to their home region or extend into third-country markets.

### 5.3 Universities and research-oriented institutions

The third step of the ATH analysis addresses the role of universities and research institutions with regard to AI development in China. Table 4 displays the initiatives of the top research institutions in the period 2017-18.

Table 4 – Main AI initiatives in China in top universities and research institutions (2017-18). Source: own compilation.

| Institution | Main initiatives in AI | Area / Fields |
|---|---|---|
| *Tsinghua University* | • State Key Laboratory of Intelligent Technology and Systems | Computer science – general research |
| *Peking University* | • Department of Machine Intelligence (formerly Center for Information Science) | Interdisciplinary research (ten departments, including mathematics, computer science, and electronic engineering)<br>Machine perception, intelligent information processing, and machine learning |
| *Zhejiang University* | • College of Computer Science and Technology | Computer science and technology – general research |
| *Shanghai Jiaotong University* | • Intelligent voice technology – computer science department<br>• School of Artificial Intelligence | Computer science – intelligent voice technology<br>Fundamental theory and technology, chips and system architecture, and applications such as smart cars |
| *University of Science and Technology of China* | • Huangpu (Whampoa) Military Academy | Industrial automation, intelligent equipment control, pattern recognition, and intelligent information processing |
| *Xi'an Electronic and Science University* | • School of AI | Electronics and science – general research |
| *Nanjing University* | • Artificial Intelligence School<br>• Nanjing Branch of JD Artificial Intelligence Research Institute | Machine learning, data mining, pattern recognition, information retrieval; evolutionary computation, neural computing and their applications |
| *Changchun University of Science and Technology* | • Artificial Intelligence School | Cognition and reasoning, machine learning; natural language understanding, computer vision, robotics, game and ethics<br>Interdisciplinary research (e.g. Photoelectric technology, |
| | | precision manufacturing, computer science, information communication, quantum and nanotechnology) |
| *Wuhan University* | • AI joint laboratory (Xiaomi Inc. and Wuhan University) | Machine translation, emotion recognition and calculation, image enhancement and noise reduction, image in depth information calculation, task-based conversation context natural dialogue generation, speech processing, intelligent question and answer |
| *Chinese Association for Artificial Intelligence (CAAI)* | • The CAAI is the only state-level science and technology organization in the field of AI in China | Unite artificial intelligence science and technology professionals – general research |
| *National Key Laboratories* | • National Engineering Laboratory for Speech and Language Information Processing (NEL-SLIP)<br>• National Engineering Laboratory of Virtual Reality (VR)/Augmented Reality (AR) Technology and Application<br>• National Engineering Laboratory of Brain-Like Technology and Application | Computer science and technology – general research |
| *Sinovation* | • Privately run research centre on AI | Upgrade thousands of traditional companies across China using AI, in particular the big state-owned enterprises – general research |

As a summary, universities and research-related institutions play two pivotal roles in AI development: (i) the provision of talent, and (ii) and the production of knowledge, which are achieved not only by training new scientific and technological talent but also by conducting basic research. This means moving beyond the focus on increasing the absolute number of students and experts, and placing greater emphasis on the quality of education (Zhang Jun, 2018).

This new strategic focus is reflected in the shift in the training and research within the AI field. Previously, AI teaching and research were carried out through the cooperation between the faculties/schools of the related majors (such as computer science, mathematics, and electronic engineering). But in April 2018,[18] the Ministry of Education issued the 'AI Innovation Action Plan for Colleges and Universities'. It proposes to establish one hundred 'AI+X' specialization categories, publish fifty world-class teaching materials for undergraduate and graduate studies, develop fifty national-level high-quality online open courses, and establish fifty artificial intelligence faculties, research institutions, or interdisciplinary research centres by the year 2020. The plan also suggests introducing education for AI at the level of primary and secondary schools (already going on at selected locations), although experts disagree about its need and real impact.

The result is that already a range of specialized graduate and postgraduate schools of AI has been founded, and the main Chinese universities are launching educational programmes and research opportunities in AI.

Regarding research, the main organization to launch calls to fund research is the National Natural Science Foundation of China (NSFC). As of 2017-18, their calls in AI were mainly oriented towards three aspects: (i) frontier research on artificial intelligence; (ii) the intelligent autonomous body; and (iii) intelligent decision making of complex processes.

As an additional relevant insight, China's academic institutions aim to increase their international exposition up (Chen & Li, 2019). Therefore, although China's AI innovation ecosystem is developing relatively independently from that in other countries, researchers based at universities and research centres in different countries are the main means of connection with other countries' ecosystems -mainly with the US- through a virtuous circle of sending excellent students to the most prestigious institutions and then promoting the return to China with attractive packages. [19]

[18] Available at http://www.moe.gov.cn/srcsite/A16/s7062/201804/t20180410_332722.html

The overall new role of universities and research-related institutions has resulted in China rising and already leading in AI rankings related to human capital and knowledge production: publications in main journals and conferences, patents, and absolute numbers of new graduates in AI, as discussed below.

### 5.4 Triple helix dynamics

Based on experts' opinions, accompanied by key insights from the desk research, this section examines the main aspects related to the interactions and flows of the agents in the ATH. These relationships are categorized in four types of flows: venture capital, human capital, knowledge production and data, which is aligned with previous approaches emphasizing the exchanges among the agents of the network as drivers of value not being these only material (goods, services) but also intangibles like knowledge or benefits (Allee, 2011). These categories are used also to describe the case studies that represent the dynamics of the AI innovation ecosystem in China in the next section.

#### *5.4.1. Venture capital*

Starting with financial capabilities, there is a wide consensus of experts acknowledging the apparently 'unlimited' venture capital and government funding support as one of the main assets for the development of AI in China. In fact, as shown by the figures of market analysts, venture capital in China has set a record high in terms of investments year by year with the consolidation of the major role of public-based investment and a deep focus on AI (KPMG Enterprise, 2019; PwC & CB Insights, 2019). There are some potential side effects, mainly due to the risk of 'not smart enough investments', crowding-out more professional investors, creating signs of a financial bubble (Lucas, 2018), and sending expert (private) venture funds to later capital rounds (series B, C, …) where it is easier to distinguish the real potential of a start-up company (Huang, Pagano, & Panizza, 2017). In fact, as a consequence, during 2019, although outside the time scope of the analysis of this paper, it is worthy to note that the amount of AI funding has decreased significantly with more stringent criteria to evaluate the viability of AI-related start-ups.

#### *5.4.2. Human capital*

Regarding the human capital capabilities, several experts have acknowledged the quantity and the quality of AI talent as one of the main challenges ahead for the development of the Chinese ecosystem. China has dramatically expanded its technical workforce, both domestically and by attracting foreign talent, but it is still behind the EU and the US not only in terms of the absolute number of experts but also in terms of their level of expertise (Elsevier, 2018; JFGagne, 2018; Ng, 2018).

One of the main questions is how to achieve a large enough provision of AI talent and ensure that the local education system is able to supply well-qualified employees, as the virtuous circle of high school education in China, education at prestigious foreign universities, and later recruitment for a high-end expert talent position in China might not suffice.[20] The difficulties of increasing the pool of talent are also related to the knowledge and skill set needed to succeed in the application of AI to a particular domain and the shortcomings of the education system, which is none too prone to innovation/creativity. As one of the experts summarized, 'there is a need of ground-breaking thinking in the underlying theory'. In this sense, 'more opening up' is viewed by the experts as a possible solution.

Similarly, a related issue from experts' survey refers to the gap between available talent and its application to existing industry, including the non-alignment between research call themes and practical implementations in (traditional) industry. This appears to be particularly important with respect to fundamental AI research, algorithms and infrastructure.

[19] In 2008, China launched the Thousand Talent Program to recruit global experts in science in technology. As of 2018, the program brought more than 7,000 experts, mostly overseas Chinese, with about 100 of them top researchers in AI.

[20] Including the possible existence of barriers to the movement of high-tech experts.

Also, as previously highlighted, this lack of a sufficient skilled workforce involves a national competition between cities/regions for talent and investments. The accumulation of talent, funds, and opportunities in new tech industries lie mainly in the East, Beijing, Shanghai, Shenzhen, and Hangzhou as the ideal cities in which to locate an AI start-up from the experts' point of view, with the unique exception of Chengdu.

#### *5.4.3. Knowledge production*

Looking at knowledge production, the lack of a common definition of AI and the difficulty of establishing clear boundaries within the AI disciplines are the main challenges to measuring the potential of AI. To cope with these limitations, conclusions combine the survey of experts with data from the most recent research available in addition to a specific analysis conducted by the authors which examines the Chinese research production in top AI journals,[21] checking AI subdomains, and comparing it with US- and EU-affiliated research in the same period. Table 5 presents the results of the classification of 1,408 papers from the top journals in AI in 2017, considering the main AI subdomains and the authors' affiliation. As a summary, China leads in the general classification and in every subdomain, with the exceptions of computer vision and perception and natural language processing and semantics – although the data here are limited. This is the same conclusion from studies done by the White House (2016), and also according to Elsevier (2018) and WIPO (2019). Looking into the content of the publications, China-based academic papers still showcase incremental innovations and not breakthrough ideas that are still coming from other geographies such as Europe and the US (Shoham et al., 2018). Going through the citations and reputation of the journals where the articles have been published, China still lags behind the US and Europe respectively, but it also shows a growth during the last two decades (Elsevier, 2018). Therefore, in summary the reach of Chinese research is still regional, but aspiring to lead globally.

Table 5 - Analysis of papers from top AI journals in 2017. Source: own analysis.

| AI subdomain | China | EU | US | No. of Papers |
|---|---|---|---|---|
| **Evolutionary and genetic computing** | 49% | 35% | 16% | 129 |
| **Neural networks** | 57% | 17% | 15% | 338 |
| **Computer vision/perception** | 35% | 35% | 30% | 418 |
| **Automated reasoning/reasoning with uncertainty/optimization** | 52% | 29% | 14% | 219 |
| **Robotics/planning/multiagent systems** | 65% | 19% | 14% | 107 |
| **Natural language processing/semantics** | 29% | 52% | 19% | 21 |
| **Machine learning/other learning techniques** | 45% | 30% | 29% | 201 |
| **Total** | **49%** | **28%** | **22%** | **1408** |

As in the case of scientific papers, China has become a global leader in terms of the number of patents (Van Roy, Vertesy & Damioli, 2019). In the specific case of AI, the US and China are largely leading in patent applications (WIPO, 2019). Like the case of academic papers, China´s position falls when considering not only the quantity but also the quality of the AI patents.[22] Among other factors, it can be mentioned that the

[21] The list has been extracted by number of citations from the 2016 edition of the *Journal Citation Report*. The list of selected journals is as follows: *IEEE Transactions on Evolutionary Computation*, *IEEE Transactions on Pattern Analysis and Machine Intelligence*, *International Journal on Computer Vision*, *IEEE Transactions on Fuzzy Systems*, *IEEE Transactions on Cybernetics*, *IEEE Computational Intelligence Magazine*, *International Journal on Neural Systems*, *IEEE Transactions on Neural Networks and Learning Systems*, *Information Fusion*, *Neural Networks*, *Integrated Computer-Aided Engineering*, *Journal of Machine Learning Research*, *IEEE Transactions on Image Processing*, *Artificial Intelligence*, *Pattern Recognition*, *Knowledge-Based Systems*, and *Medical Image Analysis*.

[22] This classification is established considering whether the patents have been granted, cited, or belong to extended patent families. According to WIPO definitions, the three criteria are defined as follows:

- Families with at least one patent granted: This is considered 'as a validation by independent patent examiners of the novelty and inventiveness of the invention for which patent protection is being sought'.

high proportion of applications filed only in China could be due to the main interest of Chinese researchers in the domestic market, while the low percentage of citations and grants could be due to the relative novelty of Chinese patents with respect to those in other countries.

In this respect, the experts corroborated these assertions. The majority of them agreed that China is already a scientific and technical superpower but also emphasized relevant limitations of the AI ecosystem, such as the 'lack of researchers for fundamental theory', 'the limited scope of research', and 'the only focus on some aspects such as facial recognition'.

#### *5.4.4. Data availability framework*

Another relevant aspect remarked on by the experts is related to the data available in China. More data is available than anywhere else due to China's huge population and its intensive usage of many types of systems and applications. In China, there are also fewer obstacles to personal data collection and usage than in other jurisdictions, and there is also a young market eager for novelties and nominally less worried about privacy. Last but not least, the government has an interest in the deployment of large AI-based applications with a social impact. However, experts emphasized that the data available in China are 'not very diverse' and are of 'relatively low quality', mainly due to the inward orientation. Therefore, some AI solutions would only be valid for niche markets and would be difficult to extend to other geographies.

As a distinctive feature of the AI ecosystem in China, the experts also called attention to the fact that the critical views regarding AI, although on the increase, are still relatively low key, almost always referring to companies' usage of AI and taking good care to not incur disloyalty towards a policy that is highly supportive of AI. While the calls for policy support voiced in the EU and the US focus on algorithmic transparency and accountability (Garfinkel, Matthews, Shapiro, & Smith, 2017) and warn about moral responsibilities outsourced to algorithms, in China, the focus is more on placing AI within the scope of the rule of law. However, the pervasive usage of AI in China has highlighted that the absence of a more forward-looking perspective could lead to a lack of verifiable algorithms or trustworthy devices and systems. As one of the experts argued, compared with the West, 'there is (in China) a greater focus on short term goals than on long term implications'.

### 5.5. ATH results: case studies within the AI ecosystem

Following the ATH model, the last step of the analysis displays how the helices interact dynamically with each other and provide specific solutions. For this, the most relevant AI projects implemented during 2017-18 have been compiled as shown in Table 6, classified by industry and key technological domain within AI. As previously mentioned, the interactions produce relationships of different types (venture capital, human capital, knowledge production, data availability framework), as discussed below and result in specific projects and solutions.

The first observation is how the overall development of these projects mimics the proposed ATH model. The top-down framework model consists of China's AI plan for leveraging the momentum provided from the government to move beyond the outcome that the market alone could provide. Next, the projects are developed rather practically from companies, based on a trial-and-error approach that allows room for experimentation and to test and refine new policies. Depending on the success of these projects, they are either discarded, modified or scaled up. In a final stage, governments adopt a more traditional role of regulators - mostly to ensure social goals- in addition to their practical support to companies. The cases of the Toutiao news service from Bytedance and Kuaishow video network backed by Tencent are paradigmatic. The huge success of Toutiao's curated news recommended by a machine learning algorithm was followed by the public discovery that this was taking the users into extreme types of contents. In April 2018, China's State Administration of Radio and Television and Cyberspace Administration asked both news website Toutiao and

- Citation of patent families: If the patent is mentioned at least 20 times in later published patents, 'this indicates the impact of the invention on later inventions'.
- Extension of patents to several patent offices: This is an indication of 'both the desire of the patent applicant to commercialize the invention in multiple markets and the market size'.

livestreaming website Kuaishou to remove obscene and violent content. The explanation from providers was that AI-based algorithms were responsible for promoting videos according to preference and popularity, regardless of the content. As a result, the services were discontinued for about one month and reformed to include now both AI and humans to censor contents and ensure mainstreaming.

A next immediate result from an overarching examination of these cases is that AI applications are widely available and have an impact on the daily lives of China's citizens, combining convenience, efficiency, personalization, and surveillance, and providing a head start for companies involved in them.

A third result is how most of these cases require the collaboration of different levels of government and companies, on one hand validating the asymmetric approach to the triple helix model as stated above, and on the other the alignment of public and private interests. Thus, governments at national and/or regional level play a prominent role defining the strategy, providing a supportive environment and in the majority of the cases being an end customer of the solution. Also, most of the cases refer to B2C provision models, and therefore it can be concluded that the alignment between government and companies' interest takes place when mass markets with individual consumers are addressed.

As a main piece of evidence, the Ministry of Science and Technology announced the establishment of the association for the new generation of artificial intelligence development planning promotion on November 2017. There, the first list of national artificial intelligence open innovation platforms has been published, which includes automatic driving relying on the Baidu platform[23], a City Brain (smart cities) relying on the Aliyun (Alibaba cloud) platform[24], medical imaging relying in the Tencent platform[25], and intelligent voice applications relying on the iFlytek platform. All of them are in collaboration with different universities and involve the participation of start-ups and companies in their respective fields. Another practical consequence is this alignment is the prevalence of AI-based solutions and projects implemented in the field of security and public safety.

As another main example, Alibaba is championing AI-based reputation systems that can be a blueprint for China's authorities' interest in using technology for a 'social credit system', or more precisely, a 'social trust system'. Most information in this system -called Zhima Credit- comes from Ant Financial's Alipay, with more than 520 million users, and automatically determines a person's creditworthiness by examining his or her daily financial transactions and connections. The system also has access to data external to the applications and derived from interactions of the user with public services and bodies. There are many other implementations of similar social trust systems from different governmental bodies at city or provincial level.

On the security and public safety sphere, there are many solutions that rely on data provided by governments. For instance, projects from Megvii / Face++ use video identification systems that have access to the data on criminals or other people under search. iFlytek provides technology to record court hearings and aid judges in the review of criminal cases of homicide, theft, telecom fraud and illegal fundraising. LLVision is a startup that uses AI-enabled glasses[26] to help authorities to spot criminals at public venues and it is already in use in railway stations for ID verification. The AI company Grisdum has teamed up with the Beijing Intellectual Property Court to develop remote trials service on WeChat. Intellifusion has worked with municipal police forces across the country to install surveillance systems in subways and streets. Besides, its AI and facial recognition technology is used by Shenzhen city council to fine jaywalkers. Similarly, Yitu Technology has provided technological support for identification of suspects at main events. Its solution called Dragonfly Eye, is implemented by about 200 local and regional policy forces, including Qingdao, Suzhou, Xiamen public transportation or Shanghai metro system. The company has also worked with Chinese customs and immigration authorities to identify wanted suspects and smugglers. Summing up, facial recognition devices are being used by the police, insurance companies, education providers or even judges, not only for personal identification but to track facial expression and determine the customer or subject attitude.

[23] In fact, the Chinese government in March 2018 gave Baidu permission to test cars on public roads in the suburbs of Beijing, making it first on the roads in China. The company's goal is to test the system in buses made by Chinese manufacturer King Long later in 2019. Baidu's initial self-driving cars will be developed with China's Chery Automobile Co, although it also has a 2021 target to produce Level 4 autonomous cars in partnership with Chinese automaker BAIC Group (Welch & Behrmann, 2018)

[24] A flagship project of Alibaba's cloud AI tools is a suite called City Brain, designed for tasks like managing traffic data and analyzing footage from city video cameras with the ultimate goal of creating a computerized, automated public services infrastructure. According to Alibaba, its implementation in its home base of Hangzhou has increased the average speed of traffic in the city by 15%. It has also been deployed in Guangzhou and Suzhou.

[25] Tencent flagship project is the ecosystem the company is building around healthcare with the help of AI. In 2014, Tencent launched WeChat Intelligent Healthcare. The platform allows users to book appointments and make payments at hospitals through WeChat accounts. This is complemented by WeSure, a medical insurance service. In 2017, Tencent launched the AI Medical Innovation System or AIMIS, an AI-powered diagnostic medical imaging service. From 2018 it is being deployed to about 100 hospitals around China. The AIMIS team then worked closely with Guangzhou's Sun Yatsen University Cancer Research Institute and used anonymized patient data to train the diagnostic component of the AI.

Another relevant example on these helices interactions is the involvement of publicly controlled companies in the development of AI projects. One of the most relevant examples are the regional/municipal transport systems, being the Shanghai or Beijing Metro a paradigm.

The innovative purchases both by local and national governments also show how the helices are related. The flagship project of iFlytek is one of the best examples. It basically consists of an AI-enabled general practitioner medical robot that passed the written test of China's national medical licensing examination in November 2017. The robot can automatically capture and analyze patient information and make an initial diagnosis. The project was officially launched in March 2018 with a pilot project with Anhui (Hefei) Provincial Hospital and the system was developed in partnership with Tsinghua University. In fact, iFlytek, launched by alumni of the University of Science and Technology of China, has founded together with the university a national key laboratory: 'National Engineering Laboratory for Speech and Language Information Processing'.

Other interesting evidence of the interactions is the funding of AI projects. Capital from public bodies, venture capitalists and big companies has been pouring into AI-related start-ups. An example is a part of Tencent´s flagship project of building an ecosystem around healthcare,[27] the company is investing in local and international health and medical start-ups, from wearable tech to genomics. The last step in this strategy is Tencent Doctorwork, a joint venture with GAW Capital, Medlinker and Sequoia China, which operates offline medical facilities called Tencent Clinics, planning to open new clinics in China during 2018 to provide offline services linked to a mobile app monitoring customers' health status.

Two additional evidences from case studies are CloudWalk, Megvii / Face++ and SenseTime. CloudWalk received a $300 million investment from the Guangzhou Municipal Government in 2017. The company is devoted to facial recognition technology and its solution is deployed across several banks, including Agricultural Bank of China, Construction Bank and Merchants Bank, for authentication at ATMs. Megvii / Face++ received $460 million in its November 2017 round of venture capital -the largest to a computer vision startup in 2017–. It was led by a China government's venture capital fund with participation from the Russian government and backed by Ant Financial (Alibaba) as well.

It is also worth mentioning the relationships and interactions of the three helices around AI training and research. An interesting evidence of these dynamics are the 'national laboratories'[28], three of the special

[26] One challenge for facial recognition software is that it struggles when running on CCTV cameras, because the picture is blurry. The wearable glasses with AI on the front end can provide immediate and more accurate feedback.

[27] An estimate says Tencent has invested about €3 billion in medical start-ups since 2014 (Lew, 2018)

[28] The three national laboratories are:

- The National Engineering Laboratory of Virtual Reality (VR) / Augmented Reality (AR) Technology and Application. It is led by Beihang University although it also has a branch at Beijing University of Technology (BIT). The organizations in collaboration are: Beijing Institute of Technology, China Electronics Standardization Institute, Goertek Inc, and Wisesoft.
- The National Engineering Laboratory of Brain-like Technology and Application. It is led by and located in University of Science and Technology of China, Hefei, Anhui. The organizations in collaboration are: Fudan University, Shenyang Institute of projects developed related to AI as part of the 'Internet +' initiative. In these projects take part different universities, companies, government dependent organizations and the initiative is led by the National Development and Reform Commission, a key governmental agency under the State Council. Each of them is focused on key innovative capabilities within the AI field

Finally, almost every leading company is collaborating with universities across the country in a range of ways. As non-exclusive examples, JD.com announced the Nanjing Branch of JD Artificial Intelligence Research Institute, which will be built near the NJU and will become the training station for the students of its Institute of Artificial Intelligence[29]. Xiaomi Inc. and Wuhan University announced the establishment of an AI joint laboratory[30]. And iFlytek[31], Tencent[32], Baidu[33], Megvii[34] or Deepblue[35] have established AI colleges in partnership with top universities.

Table 6 – Relevant ongoing AI company case studies in China (2017-18). Source: own compilation.

| Organization – case study | Industry | Key technology | Description | Agents (Helix) involved | Relationship (s) |
|---|---|---|---|---|---|
| Alibaba – City brain | Government | Cloud | City management | • Government (national and regional/local level<br>• Industry ( large companies) | • Venture capital<br>• Knowledge production<br>• Data availability framework |
| Alibaba – DoctorYou | Health care | Image recognition | Medical diagnosis | • Government (national and regional/local level<br>• Industry ( large companies) | • Venture capital<br>• Knowledge production<br>• Data availability framework |
| Alibaba – Pig | Agriculture | Image recognition | Pig rearing | • Industry ( large companies) | • Data availability framework |
| Alibaba – Supermarkets | Retail | Face recognition | Supermarkets without employees | • Government (national and regional/local level<br>• Industry ( | • Venture capital<br>• Data availability |

Automation Chinese Academy of Sciences, Institute of Microelectronics of Chinese Academy of Sciences, Institute of Neuroscience of Chinese Academy of Sciences, Baidu Inc., Microsoft Research Asia, and iFlytek.

- National Engineering Laboratory of Deep Learning Technology and Application. It is led by and located in Baidu Inc. The organizations in collaboration are: Tsinghua University, Beihang University, China Electronics Standardization Institute, and China Academy of Information and Communications Technology.

[29] https://www.nju.edu.cn/_t465/f7/d4/c3814a260052/page.htm

[30] http://fonow.com/view/206675.html

[31] Partnership with Southwest University of Political Science AI Law School (Chongqing), the Chongqing University of Posts and Telecommunications, Nanning University (Guangxi), Anhui University School of Information Technology, Jiangxi Vocational College of Applied Technology and Chongqing Vocational College of Science and Technology

[32] Partnership with Shenzhen University, Liaoning Technical University, Shandong University of Science and Technology and Liaocheng University (in Shandong)

[33] Partnership with Jilin University, Hunan Normal University, Henan Institute of Finance

[34] Partnership with Nanjing University and Xi'an Jiaotong University

[35] Partnership with Central South University and Jiangsu University of Technology of Technology

| | | | | | |
|---|---|---|---|---|---|
| | | | | large companies and start-ups) | framework |
| Alibaba – Zhima | Ecommerce | Online transactions analysis | Rate customer reputation – social credit | • Government (national and/or regional/local level<br>• Industry (start-ups and large companies) | • Venture capital<br>• Human capital<br>• Knowledge production<br>• Data availability framework |
| Baidu | Automobile | Driverless car | 2019 for highways | • Government (national and/or regional/local level<br>• Industry (start-ups / large companies) | • Venture capital<br>• Human capital<br>• Knowledge production<br>• Data availability framework |
| Baidu – DeepVoice | Media | Text-to-speech | Mimic any voice | • Government (national )<br>• Industry (start-ups / large companies)<br>• Universities & Research centres | • Venture capital<br>• Human capital<br>• Knowledge production<br>• Data availability framework |
| Baidu + Huawei | Telecoms | Open platform for mobile devices | Mobile app developers | • Industry (/ large companies) | • Venture capital<br>• Human capital<br>• Knowledge production<br>• Data availability framework |
| Beijing University | Education | Face recognition | Enter into dorms | • Government (national and/or regional/local level<br>• Universities & Research centres | • Venture capital<br>• Human capital<br>• Knowledge production<br>• Data availability framework |
| ByteDance | News / Social networks | Personalized content | Curated news and videos | • Industry (start-ups) | • Data availability framework |
| CloudWalk | Banking | Face recognition | ATM operations | • Government (regional/local level<br>• Industry (start-ups / large companies) | • Venture capital<br>• Data availability framework |
| Chinese Academy of | Government | Data mining | Diplomacy | • Government (national and/or | • Venture capital |
| Sciences | | | | regional/local level<br>• Universities & Research centres | • Human capital<br>• Knowledge production<br>• Data availability framework |
| Didi Chuxing | Transportation | Driverless car | Self-driving taxi service | • Government (national and/or regional/local level<br>• Industry (start-ups / large companies)<br>• Universities & Research centres | • Venture capital<br>• Human capital<br>• Knowledge production<br>• Data availability framework |
| Megvii / Face++ | Security | Face recognition | Gov. records data on criminals/missing people | • Government (national and/or regional/local level<br>• Industry (start-ups / large companies)<br>• Universities & Research centres | • Venture capital<br>• Human capital<br>• Knowledge production<br>• Data availability framework |
| Gridsum | Government | Document management | Remote trials in intellectual property cases | • Government (national level)<br>• Industry (start-ups / large companies) | • Venture capital<br>• Knowledge production<br>• Data availability framework |
| Heaven's Temple | Tourism | Face recognition | Retrieve toilet paper | • Government (national and/or regional/local level | • Data availability framework |
| iFlytek | Health care | Robot/user interface | Patient diagnosis | • Government (national and/or regional/local level<br>• Industry (start-ups / large companies)<br>• Universities & Research centres | • Human capital<br>• Knowledge production<br>• Data availability framework |
| iFlytek | Law | Voice recognition | Criminal cases hearings | • Government (national and/or regional/local level<br>• Industry (start-ups / large companies) | • Venture capital<br>• Data availability framework |
| Intellifusion | Security | Surveillance | Subways and streets. Taxi drivers' identification. Jaywalkers | • Government (national and/or regional/local level<br>• Industry (start- | • Data availability framework |
| | | | | ups / large companies) | |
| KFC | Food | Face recognition | Menu acquisition and dietary recommendations | • Industry (start-ups /large companies) | • Data availability framework |
| LLVision | Security | Face recognition | Verification of ID at public venues through glasses | • Government (national and/or regional/local level<br>• Industry (start-ups / large companies) | • Knowledge production<br>• Data availability framework |
| NIO | Automobile | Voice recognition | 2018 mass production | • Government (national and/or regional/local level<br>• Industry (start-ups / large companies)<br>• Universities & Research centres | • Venture capital<br>• Human capital<br>• Knowledge production<br>• Data availability framework |
| Ping An | Insurance | Facial expressions | Insurance premium if the customer tells the truth | • Government (national and/or )<br>• Industry (start-ups / large companies)<br>• Universities & Research centres | • Venture capital<br>• Human capital<br>• Knowledge production<br>• Data availability framework |
| Policy Cloud | Security | Information integration | Trace suspects and anticipate behaviours | • Government (national and/or regional/local level<br>• Industry (start-ups / large companies)<br>• Universities & Research centres | • Knowledge production<br>• Data availability framework |
| SenseTime | Security | Face recognition | Surveillance footage | • Government (national and/or regional/local level<br>• Industry (start-ups / large companies)<br>• Universities & Research centres | • Venture capital<br>• Human capital<br>• Knowledge production<br>• Data availability framework |
| Shanghai Marathon | Leisure | Face recognition | Personalized video of the race | • Government (national and/or regional/local level<br>• Industry (start-ups / large companies)<br>• Universities & Research centres | • Data availability framework |
| Shanghai Metro | Transportation | Face/voice recognition | Tickets, recommended route | • Government (national and/or regional/local level<br>• Industry (start-ups / large companies) | • Venture capital<br>• Data availability framework |
| Shenzhen – Hong Kong Border | Security | Face recognition | Automatic border crossing | • Government (regional/local level<br>• Industry (start-ups /)<br>• | • Data availability framework |
| Tencent – AIMIS | Health care | Image recognition | Disease diagnosis | • Government (national and/or regional/local level<br>• Industry (start-ups / large companies)<br>• Universities & Research centres | • Venture capital<br>• Human capital<br>• Knowledge production<br>• Data availability framework |
| Tencent – Wechat | Governance | Face recognition | Digital ID | • Government (national )<br>• Industry (start-ups / large companies)<br>• Universities & Research centres | • Venture capital<br>• Knowledge production<br>• Data availability framework |
| Xio Liange | Security | Surveillance | Gov. project to monitor people in 55 cities | • Government (national and/or regional/local level<br>• start-upsUniversities & Research centres | • Knowledge production<br>• Data availability framework |
| Yitu – Dragonfly Eye | Security | Face recognition | Customs, immigration identification, public transportation | • Government (national and/or regional/local level<br>• Industry (start-ups / large companies)<br>• Universities & Research centres | • Knowledge production<br>• Data availability framework |
| Yitu – AICare | Health care | Image recognition | Medical test analysis | • Government (national and/or regional/local level<br>• Industry (start-ups / large companies)<br>• | • Knowledge production<br>• Data availability framework |
| 17zuoye | Education | Personalized content | Learning multi-platform (students, teachers, parents) | • Industry (start-ups / large companies) | • Venture Capital |

## 6. Conclusions

Drawing from the innovation ecosystem theory, this paper proposes an asymmetric Triple Helix (ATH) model adapted to the characterization of the AI innovation ecosystem in China. It includes a detailed qualitative examination of the ecosystem in China from the point of view of the three TH spheres, describing how they are being built, their dynamic relationships and the results so far. On authors' view, this is a useful approach to extract of insights that are not directly available through quantitative analysis when analysing innovative industries in a particular geography.

The bottom line of the existing successful implementation of AI solutions in China consists of the three components of the TH aligned on top of a very unique and favourable environment. The Government at different levels has contributed with the conditions -plans, strategies, regulations, room for experimentation- and the practical support -venture capital, purchases, access to data- for AI innovation to happen, favouring (national) security, stability, positive impact on social welfare and a strict control of cyberspace over privacy and individual rights. China's technology industry, led by Internet giants Alibaba, Tencent and Baidu, are building research centres, deploying applications and hiring all the available AI talent. Besides, a middle-class of AI-based companies with high impact applications is burgeoning. In addition, increasingly excellent universities are launching educational programmes and research opportunities in AI and China performance in terms of AI patents and publications is improving significantly.

The overall result is a strong alignment of companies and government interests as a whole with the ecosystem most developed in the areas of image, face and voice recognition, together with their supporting techniques, such as machine and deep learning, as revealed by the analysis. Consequently, the most developed AI application domains in China are security, healthcare, transportation, traffic management —from the perspective of cities—, commerce, payments and driverless cars, all of them belonging to the B2C domain, and most of them with a public slant.

The analysis also shows the main gaps of the ecosystem in the shape of low adoption of AI within traditional industries, patchy implementations, the gap between the quantity and the quality of the pool of AI talent, the dependence on foreign industries, and the relatively low quality and diversity of the data. In addition, AI presents practical risks and ethical issues in terms of social governance.

Particular notes of interest from the analysis are the particular approach to policy making, and the clustering of the innovation ecosystem. In fact, the approach to policy making is radically practical, based on several different -sometimes hardly compatible- strategies, all of them based on a trial-and-error approach by testing new policies in starting projects. Depending on the outcomes of these tests, AI policies and strategies are either abandoned, refined or rolled out across the country. Also because of the head start of these experiments, major opportunities tend to concentrate in some regional/local specialized clusters competing for talent and investments across the country.

An additional insight from the analysis is that the ecosystem is developing relatively independently from that of other countries, with the exception of the strategic pledge for the internationalization of research institutions. TH connections outwards are made between researchers based at universities and research centres in different countries, as well as through major Chinese technology companies that have set up basic research centres and funding in the US and, vice versa, via a number of US technology companies that have established research centres in China to tap into local talent and developments. The EU's and other countries such as Japan, Russia or South Korea's presence in China's AI ecosystem is low, with some minor exceptions through companies, university collaboration and individuals working in start-ups and venture capital.

This paper is also relevant in showing how a qualitative approach could be useful to address the complexity of innovation ecosystems, especially in those cases such as China in which the availability and reliability of statistical data have sometimes been controversial (Koch-Weser, 2013). As demonstrated, the combination of document analysis and an experts' survey has been suitable for identifying the key players in the ecosystem, describing their roles, and framing their relationships in the novel and emerging context of innovation, as is the case of AI development. This qualitative knowledge could help subsequent quantitative characterizations by providing a conceptual scheme of analysis.

The results in this paper are obviously a synopsis of the extremely complex and evolving AI landscape in a unique country such as China. The focus of the paper has been on developing a theoretical framework for the innovation ecosystem in China and applying it to the AI landscape from a descriptive and qualitative perspective, but it still requires a further quantitative review to evaluate its explanatory capacity. Therefore, future research could develop the framework further to enhance its generalizability to other fields and/or geographies. In addition, future analysis could use further developments of the Triple Helix to include the interrelations of citizens (Quadruple Helix) and the environment (Quintuple Helix), identifying potential differences with respect to other geographies.

Despite these limitations in the analysis, we can conclude that the AI ecosystem in China is inward oriented and led by an alignment of the three agents of the TH. While it is true that governments play a leading role, the three helices are strongly related and their very dynamic interactions are driving the evolution of the whole ecosystem. The overall result can be summarized as China's AI plan for leveraging the momentum provided by both the Government and private companies to move beyond the outcome that the market alone – either before or after regulation – could provide. In fact, AI deployments with a high impact on daily activities have already started, including authentication processes, medical diagnosis, premium insurance, transportation and retail facilities, and security. Adding all these together configure a scenario of deep economic and social change, in which convenience, efficiency, personalization, and surveillance are inextricably combined. This is not the same situation as in the EU and the US, dominated by different combinations of regulation and market forces. This is mainly because China has been able to mobilize resources for technology sectors aimed at final consumers such as AI in a way in which democracies are simply unable to do because of the limits of government power or popular mandate, combined with a long-term vision of technological development that is independent of the policy cycle.

## Annex

**Experts Survey**

**AI innovation ecosystem's status, challenges and prospects**

1- To evaluate the current health status of the AI innovation ecosystem in China, which mark would you give?
如果评估当前中国人工智能创新生态系统的健康状况，您会给出什么分数？（1糟糕 - 5优秀）

2- Why? Please explain
请说明理由

3- Which are the prospects of the AI innovation ecosystem in China in the next 5 years?
未来5年中国人工智能创新生态系统的前景如何？（1非常消极 - 5非常有希望）

4- Why? Please explain
请说明理由

Which do you think is the biggest challenge / gap for the China AI roadmap to 2030 (aim: 'to become world leader in AI')? (open question) 您认为在中国人工智能达到2030年规划目标（'成为人工智能领域的领导者'）的最大挑战/差距是什么？（开放问题）

5- How would you address the above challenge / gap? (open question)
针对这个挑战您会有什么建议?(开放问题)

6- Which industry do you think will be more disrupted by AI in China? Why? (open question)
你认为在中国哪个行业会被人工智能影响最大？为什么呢？（开放问题）

**AI ecosystem's leading actors, advantages and disadvantages**

7- Will innovation in AI in China come primarily from large companies or from start-ups? Please explain the reason
人工智能在中国的创新主要会来自大公司还是初创公司？ 请说明原因

8- Will new AI start-ups be able to challenge existing large companies in AI in China or will they be acquired by existing Internet giants at some point? (Open question)
新的AI初创公司能够挑战AI中现有的大型公司吗？还是会在某些时候被收购？ （开放问题）

9- Which advantage do you think a Chinese company in AI has compared to a foreign company to succeed in China? (open question)
您认为中国的AI公司与外国公司相比，有哪些优势？ （开放问题）

10- What advantage do you think a foreign company in AI has compared to a Chinese company to succeed in China? (open question)
您认为外国的AI公司与中国公司相比，有哪些优势？ （开放问题）

11- Which is the role of patents in AI in China?<br>专利在中国人工智能的发展起到什么作用？ （1无关 - 5非常重要 ）

12- Which is the main advantage of venture capital for AI in China? (open question)
人工智能风险投资在中国最大的的优势是什么？（开放问题）

14- Which is the main disadvantage of venture capital for AI in China? (open question)
人工智能风险投资在中国最大的劣势是什么？（开放问题）

**AI opportunities and threats in China**

15- Which is the weakest R&D area in AI in China? (open question)
中国最薄弱的AI研究领域是什么？（开放问题）

16-Which cities (or specific areas of a city) would you locate an AI startup in China? (open question)
中国的哪个城市（或者城市的特定地区)最适合创立AI初创企业？（开放问题）

17- Which will be the role of algorithm transparency and accountability in AI in China? (from 0- no role whatsoever to 5 – key role for developments)
算法透明度和问责制将在中国AI发展中起到什么作用？ （从0-没有任何作用到5--发展的关键角色）

18- How probable is a fragmented (solutions not compatible) AI / tech scenario between regions of the world? (from 0 – no fragmentation at all to 5 – solutions fully incompatible across constituencies)
世界各地区之间存在零散化AI技术情景（解决方案不兼容）的可能性如何？ （从0 - 完全没有分散化到5 - 解决方案在各地区完全不兼容）

19- How probable is AI contributing to harm of people at large in the mid-term?
在中期来讲，人工智能有多大可能会广泛的对人们造成伤害？

**Personal background**

20- How many years of experience do you have in tech industry?
在科技行业从业年数？

21- How many years of experience do you have in AI industry?
在人工智能行业从业年数?

22- How many years of experience do you have in academia related to technology?
在科技相关学术界从业年数?

23- How many years of experience do you have in academia related to AI?
在人工智能相关学术界从业年数？

24- How many years of experience do you have in gov. institutions related to technology?
在科技相关政府机构从业年数?

25- How many years of experience do you have in gov. institutions related to AI?
在人工智能相关政府机构从业年数?

26- How many years have you worked in China?
请问您在中国的工作年限？

27- Which is your gender? / 性别

28- Which is your age? / 年龄

29- In which area of AI do you work? (if other, please indicate)*

AI technology in general / 基础人工智能技术
Ecomics / business of AI / 经济/商业
Social impact of AI / AI社会影响
Institutional approach to AI (government bodies, public funding,…) / AI的体制方法（政府，公共基)
Neural networks / 神经网络
Computer vision / 计算机视觉
Automated reasoning / 自动推理
Knowledge-based systems / 基于知识的系统
Robotics / 机器人
Natural language processing / 自然语言处理
Evolutionary and genetic computing / 进化和遗传计算
Machine learning / 机器学习
Other

30- In which application of AI do you work? (if other please, indicate)*

Agriculture / 农业
Automobile, transportation, smart driving / 汽车，交通，智能驾驶
Commerce, retail / 商业，零售
Consumer electronics / 消费类电子产品
Education / 教育
Energy / 能源
Food / 食品
Finance / 金融
Healthcare / 健康
Insurance / 保险
Legal / 法律
Manufacturing / 制造业
Marketing / 市场
Media / 媒体
Real state / 房地产
Security / defence / 安全/保卫
Sports / 体育
Telecoms / 电信
Tourism / 旅游
Other

31- Which is the best description of your position in the company/organization?
请问如何描述您在公司的职位？

Senior Management / 高管
Manager / 经理
Software developer / 软件工程师
Hardware developer / 硬件工程师
Algorithm engineer / 算法工程师
Product Manager / 产品经理
Professor / 教授

Researcher / 研究人员
Other

32- Which is the best description of your company / institution? (if other please, indicate)
请问如何描述您所在公司?

Governmental body or institution / 政府机构
Large – listed company / 大型公司
Small / medium Enterprise / 中小企业
Start-up / 初创企业
Venture capital – Business angel / Series A / 风险投资-天使/A 轮
Venture capital – Series B and beyond / 风险投资-B 轮或以上
University / 大学
Research centre / 研究中心
Other

Thank you for your participation
谢谢你的参与